\documentclass[preprint,aip,jcp]{revtex4-1}
\usepackage{epsfig}
\usepackage{textcomp}
\usepackage{graphicx}
\usepackage{amsmath}
\usepackage{url}
\usepackage[normalem]{ulem}

\usepackage{color}
\usepackage{xr}
\definecolor{Myblue}{rgb}{0.3,0.3,1.0}

\usepackage{rotating}

\begin{document}
\title{Thermal Activation of Methane by MgO$^+$: Temperature Dependent
  Kinetics, Reactive Molecular Dynamics Simulations and Statistical
  Modeling}

\author{Brendan C. Sweeny} \affiliation{NRC postdoc at Air Force
  Research Laboratory, Space Vehicles Directorate, Kirtland Air Force
  Base, New Mexico 87117}

\author{Hanqing Pan} \affiliation{USRA Space Scholar at Air Force
  Research Laboratory, Space Vehicles Directorate, Kirtland Air Force
  Base, New Mexico 87117}

\author{Asmaa Kassem} \affiliation{USRA Space Scholar at Air Force
  Research Laboratory, Space Vehicles Directorate, Kirtland Air Force
  Base, New Mexico 87117}

\author{Jordan C Sawyer} \affiliation{NRC postdoc at Air Force
  Research Laboratory, Space Vehicles Directorate, Kirtland Air Force
  Base, New Mexico 87117}

\author{Shaun G. Ard} \email[]{rvborgmailbox@us.af.mil}
\affiliation{Air Force Research Laboratory, Space Vehicles
  Directorate, Kirtland Air Force Base, New Mexico 87117}

\author{Nicholas S. Shuman} \affiliation{Air Force Research
  Laboratory, Space Vehicles Directorate, Kirtland Air Force Base, New
  Mexico 87117}

\author{Albert A. Viggiano} \affiliation{Air Force Research
  Laboratory, Space Vehicles Directorate, Kirtland Air Force Base, New
  Mexico 87117}

\author{Sebastian Brickel} \affiliation{Department of Chemistry,
  University of Basel, Klingelbergstrasse 80, CH-4056 Basel,
  Switzerland\\ Present Address: Department of Chemistry - BMC,
Uppsala University, BMC Box 576, 751 23 Uppsala,Sweden}

\author{Oliver T. Unke} \affiliation{Department of Chemistry,
  University of Basel, Klingelbergstrasse 80, CH-4056 Basel,
  Switzerland\\ Present Address: Machine Learning Group, TU Berlin,
  Marchstr. 23, 10587 Berlin, Germany}

\author{Meenu Upadhyay} \affiliation{Department of Chemistry,
  University of Basel, Klingelbergstrasse 80, CH-4056 Basel,
  Switzerland}

\author{Markus Meuwly} \email[]{m.meuwly@unibas.ch}
\affiliation{Department of Chemistry, University of Basel,
  Klingelbergstrasse 80, CH-4056 Basel, Switzerland}

\date{\today}

\begin{abstract}
The kinetics of MgO$^+$ + CH$_4$ was studied experimentally using the
variable ion source, temperature adjustable selected ion flow tube
(VISTA-SIFT) apparatus from 300 $-$ 600 K and computationally by
running and analyzing reactive atomistic simulations.  Rate
coefficients and product branching fractions were determined as a
function of temperature.  The reaction proceeded with a rate of $k =
5.9 \pm 1.5 \times 10^{-10} (T / 300 $ K$)^{-0.5 \pm 0.2}$ cm$^3$
s$^{-1}$.  MgOH$^+$ was the dominant product at all temperatures, but
Mg$^+$, the co-product of oxygen-atom transfer to form methanol, was
observed with a product branching fraction of $0.08 \pm 0.03 (T / 300
$ K$)^{-0.8 \pm 0.7}$. Reactive molecular dynamics simulations using a
reactive force field, as well as a neural network trained on thousands
of structures yield rate coefficients about one order of magnitude
lower. This underestimation of the rates is traced back to the
multireference character of the transition state
[MgOCH$_4$]$^+$. Statistical modeling of the temperature-dependent
kinetics provides further insight into the reactive potential surface.
The rate limiting step was found to be consistent with a four-centered
activation of the C-H bond, consistent with previous calculations.
The product branching was modeled as a competition between
dissociation of an insertion intermediate directly after the
rate-limiting transition state, and traversing a transition state
corresponding to a methyl migration leading to a Mg-CH$_3$OH$^+$
complex, though only if this transition state is stabilized
significantly relative to the dissociated MgOH$^+$ + CH$_3$ product
channel.  An alternative non-statistical mechanism is discussed,
whereby a post-transition state bifurcation in the potential surface
could allow the reaction to proceed directly from the four-centered TS
to the Mg-CH$_3$OH$^+$ complex thereby allowing a more robust
competition between the product channels.
\end{abstract}

\maketitle

\section{Introduction}
The thermal activation of methane has long remained a ``holy grail''
in chemistry, offering the potential to transform the chemical economy
by providing abundant feedstock for multitudes of value added
chemicals.\cite{Olah2013,Douglas2013,Palkovits2010} Gas phase ion
reactivity combined with quantum chemical calculations provides the
ability to study the mechanistic underpinnings of catalyzing this
activation free from the complications inherent to bulk or solution
phases.\cite{Shiota2000,Schwarz2011,Li2016,Schwarz2015,Schroder2010}
The observation of methane activation by MgO$^+$ is of particular
note,\cite{Schroder2006} as prior studies had focused on transition
metals, such as the well performing but expensive palladium-based
catalysts.\cite{glorius:2018} Detailed understanding of methane
activation by main group metal oxide species such as MgO$^+$ presents
the potential to develop cost effective catalysts for large scale
utilization of this important chemical feedstock. \\

\noindent
Temperature-dependent kinetics offers insight into catalyst
development in several ways. Comparing this data with statistical
modeling of calculated reaction surfaces can confirm or refute a
proposed mechanism as well as help identify energetic and entropic
factors governing reactivity. We have employed this technique for
several transition metal oxides reacting with
methane,\cite{Ard2014,McDonald2018} adding insight into the key rate
limiting and product determining kinetic features of these systems.
Additionally, the variation of rate coefficient and product branching
for a given reaction as a function of temperature aids identifying the
optimum conditions for a desired reaction product. Conversely,
atomistic simulations based on accurate, reactive force fields provide
molecular-level mechanistic insight into gas-phase reactions ranging
from triatomics\cite{Ivanov2007,MM.cno:2018,Caridade2018,MM.no2:2020}
to systems such as the Diels-Alder reaction between
2,3-dibromo-1,3-butadiene and maleic anhydride.\cite{MM.da:2019} \\

\noindent
In the present work, we consider the hydrogen abstraction reaction
\begin{equation}
  ^2{\rm MgO}^+ + {^1\rm CH}_4 \rightarrow {^1\rm MgOH}^+ + {^2\rm CH}_3
\label{eq:1}
\end{equation}
associated with $\Delta_r H_{\rm 0 K} = -0.77 \pm 0.15$ eV and the
oxygen atom transfer reaction
\begin{equation}
  ^2{\rm MgO}^+ + {^1\rm CH}_4  \rightarrow {^2\rm Mg}^+ + {^1\rm CH}_3{\rm OH}
\label{eq:2}
\end{equation}
with $\Delta_r H_{\rm 0 K} = -1.34 \pm 0.10$ eV. The reaction
thermochemistry is derived from experimental determinations of the
MgO$^+$ bond dissociation energy (2.5 eV)\cite{DALLESKA1994203} and
the MgO proton affinity (988 kJ/mol).\cite{Lias,Linstrom1997}\\

\noindent
Reaction \ref{eq:1} has been previously investigated at room
temperature using an ion trap apparatus and through \textit{ab initio}
calculations\cite{Schroder2006} and computationally from density
functional theory and higher-level correlated
calculations.\cite{schwarz:2019} Although oxygen atom transfer from
the oxide cation forming methanol is energetically preferred, it was
observed for $<$ 2\% of reactions.  The stationary points were
calculated at the MP2/6-311 (2d,2p) level, and were found to be in
qualitative agreement with the observed efficiency of $ \sim $40\%.
The rate limiting step was concluded to be a four-centered transition
state leading to formation of a [CH$_3$-MgOH]$^+$ insertion complex,
as opposed to a direct hydrogen abstraction pathway which was
calculated to lie nearly isoenergetic with reactants and 0.1 eV above
the four-centered TS. Previous work\cite{Sweeny2019} on the reaction
of MnO$^+$ + CH$_4$ has highlighted the importance for entropic
considerations, as in that case a direct hydrogen abstraction was
found to be competitive with an energetically preferred four centered
transition state similar to that for the present system.\\

\noindent
In the present work the temperature dependent kinetics is measured in
an ion flow tube and compared with computational work based on
reactive molecular dynamics simulations and complementary
computational studies. This provides a comprehensive characterization
of the first step of the reaction and also highlights future
possibilities to further investigate this important process.

\section{Experimental Methods}
All measurements were performed on a recently described variable ion
source, temperature adjustable selected ion flow tube
(VISTA-SIFT).\cite{ard:2019} For these measurements a laser
vaporization (LaVa) ion source was used. MgO$^+$ ions were produced by
ablating a translating, rotating magnesium rod (99.9\%, ESPI Metals)
by $\sim$3 mJ/pulse of the second harmonic (532 nm) of an Nd:YAG
(Litron) laser operating at 100 Hz. The ablated metal was entrained in
a pulsed supersonic expansion (General Valve, Series 9) of a 60:30:10
mixture of He/N$_2$O/Ne at a backing pressure of 55 PSI with a pulse
width of 200 $\mu$s, also operated at 100 Hz.  Upon formation, ions
are transported using a series of radio frequency (RF) ion guides and
ion optics to a quadrupole mass filter.  Mass-selected ions are
injected into a 7.3 cm diameter, 1 m long flow tube via a Venturi
inlet at $\sim$0.35 Torr of helium buffer allowing approximately
10$^4$ $-$ 10$^5$ collisions for thermalization with the flow tube
walls prior to introduction of the neutral reagent.  The temperature
of the flow tube is varied from 300 to 600 K by means of resistive
heating.  The reagent (CH$_4$, 99.98\% Sigma Aldrich) is added using a
mass flow controller (MKS) 59 cm before the terminus of the flow tube
providing 2-3 ms of reaction time prior to extraction through of 4 mm
diameter aperture in a nose cone biased to $\sim$ -5 V.  The ions are
transported using an RF ion guide to the entrance of an
orthogonally-accelerated time-of-flight mass spectrometer.  Relative
abundances of reactant and product ions are monitored as a function of
reagent concentration allowing determination of rate coefficients and
product branching.  The measured rate coefficients are ascribed an
uncertainty of $k \pm 30$\%, while the product branching fraction
uncertainty depends on temperature, see Table \ref{tab:tab1}.

\section{Computational Methods}
Reactive MD simulations were run either with CHARMM,\cite{Brooks2009}
using multisurface adiabatic reactive MD (MS-ARMD)\cite{Nagy2014} or
with the atomic simulation environment (ASE)\cite{larsen2017atomic}
when the neural network-learned potential energy surface (PES, see
below) was used.

\subsection{Reactive Force Fields}
\noindent
\textit{MS-ARMD:} MS-ARMD relies on the mixing of several atomistic
force fields, one for each connectivity. In the present case, this
includes MgO$^+$ and CH$_4$ in the reactant channel and MgOH$^+$ and
CH$_3$ in the product state. The parametrized force fields (FFs) for
the reactant and product complexes were obtained by an iterative
procedure, starting with reference parameters from
SwissParam\cite{zoete.jcc.2011.swissparam} for the four molecules. The
energy at the equilibrium geometry of the product complex was chosen
as the global zero of energy.  MP2/6-311+G(2d,2p) energies for
representative structures from 250 ps dynamics were calculated in
Gaussian09.\cite{g09} These energies were the reference for the
parametrization of the FFs. A downhill simplex
algorithm\cite{Nelder1965} was used for the fitting. After
parametrizing the individual molecules the optimized FFs were combined
to determine the reactant and product complex FF, respectively. The
quality of the resulting FFs over the test and validation set is shown
in Figure~\ref{fig:fig6}A.\\

\noindent
To parametrize the adiabatic barrier, the intrinsic reaction
coordinate (IRC) of the reaction was also calculated at the
MP2/6-311+G(2d,2p) level of theory. The structures along the IRC were
extracted and their MS-ARMD energy evaluated. A genetic algorithm was
used to parametrize the GAussian $\times$ POlynomial (GAPO)
functions\cite{Nagy2014} and to reproduce the energetics along the
reaction path. While it is possible to correctly describe the
transition between the van der Waals minimum (INT1) and the TS, see
inset Figure \ref{fig:fig6}A, the energy between the reactant state
(MgO$^+$ + CH$_4$) and INT1 is considerably underestimated. Compared
with the value of --0.99 eV (--22.90 kcal/mol) at the
MP2/6-311+G(2d,2p) level of theory it is only --0.27 eV (--6.21
kcal/mol) from the MS-ARMD parametrization. Hence, instead of a
submerged barrier at the four-centered transition state (TS1) which
can be reached from the reactant state, translational or internal
energy is required for the reaction pathway:
reactant$\rightarrow$INT1$\rightarrow$TS1$\rightarrow$product. \\

\noindent
\textit{Neural Network:} As an alternative to the parametrized MS-ARMD
force field, a neural network (NN) based on the PhysNet\cite{Unke2019}
architecture was trained on energies, forces, and dipole moments (see
Ref.~\citenum{Unke2019}) of 145\,000 structures (validated on 5000
structures), randomly selected from a set of 154\,368 structures
calculated at the MP2/aug-cc-pVTZ level of theory. The correlation
between the 4368 remaining structures (test set) and the predictions
of the trained NN with $R^2 = 1-4.1 \times 10^{-6}$ is shown in
Fig.~\ref{fgr:NNfit}. The dataset was constructed starting from 36063
structures for the CH$_4$ + MgO$^+$ and CH$_3^+$ + MgOH channels
generated from simulations with the MS-ARMD force field, and then
expanded using adaptive
sampling\cite{behler2014representing,behler2015constructing} similar
to the procedure described in Ref.~\citenum{Unke2019}. The NN
constructs a descriptor vector for each atom which encodes information
about the local chemical environment of the atoms.\cite{MM.nn:2018}
The total energy of the system is obtained by combining `atomic energy
contributions' predicted from these descriptors. The NN is 10 layers
deep with 64 neurons per layer.\\

\subsection{Molecular Dynamics Simulations}
\noindent
\textit{MS-ARMD:} The MD simulations for the reactive collisions were
carried out with CHARMM\cite{Brooks2009} and started with suitable
initial conditions (500 structures). The individual molecules,
i.e. MgO$^+$ and CH$_4$, were separately minimized (ABNR 10000 steps),
heated to the simulation temperature (12.5 ps), and equilibrated (12.5
ps). After a free dynamics (12.5 ps) for each system the centers of
mass of the two molecules were positioned 13 \AA\/ apart (along the
$x$-axis) and MgO$^+$ was randomly rotated.  \\

\noindent
\textit{Neural Network:} These simulations were carried out with
ASE.\cite{larsen2017atomic} Initial conditions for the NN were
generated along the same line as for MS-ARMD by separately heating
both molecules. Momenta were assigned from a Maxwell-Boltzmann
distribution at the desired temperature followed by $NVT$ Langevin
dynamics. The time step was 0.1 fs for a total simulation time of 1 ps
(for generating the initial conditions). Then, simulations were
started by positioning the centers of mass of the collision partners
13 \AA\/ apart.\\

\noindent
\textit{Stratified Sampling:} Reactive molecular dynamics simulations
were performed using stratified sampling for
$b$.\cite{Bernstein1979,Denis-Alpizar2017} For the stratified sampling
all trajectories were grouped in non-overlapping intervals of $b$ with
a weight
\begin{equation}
V_k = \frac{(b_k^+)^2 - (b_k^-)^2}{b_{max}^2}
\end{equation}
where $b_k^+$ and $b_k^-$ are the maximum and minimum value of
$b$. The reaction probability for each interval is then given by
\begin{equation}
P_{reac} = \sum_k V_k \frac{N_k}{N_{k,tot}}
\end{equation}
where $k$ labels a given interval, and $N_k$ and $N_{k,tot}$ are the
number of reactive and total number of trajectories in that
interval. The rate coefficient is then
\begin{equation}
k(T) = g(T) \sqrt{\frac{8 k_B T}{\pi \mu}}  \pi b_{max}^2 * P_{reac}
\end{equation}
where $k_B$ is the Boltzmann constant, $T$ is the temperature, $\mu$
is the reduced mass of MgO$^+$ and CH$_4$, and $g(T) = 1$ is the ratio
of the electronic degeneracy factors for reactants and products.\\

\noindent
Simulations were performed for $b = 0$ to $b_{\rm max}$ ($\Delta b =
0.5$ \AA\/) by accelerating the two molecules towards their common
center of mass. As the MS-ARMD parametrization was unable to correctly
describe the prereactive complex (see above), the energy difference
between the MgO$^+$+CH$_4$ asymptote and the prereactive complex (18.6
kcal/mol) was included by scaling the velocities of all atoms
accordingly, which amounts to an additional kinetic energy of $\sim
2.5$ kcal/mol/atom of internal energy. This is akin to recent
explorations of the Diels-Alder reaction between maleic acid and
2,3-dibromo-1,3-butadiene.\cite{MM.da:2019} A total of 3500
simulations per temperature were run with MS-ARMD.\\

\noindent
For the simulations using the NN, two sets of simulations were run. In
one of them, 3500 trajectories per temperature for $t_{\rm fin} = 500$
ps each, with drawing impact parameters $b$ from a uniform
distribution. For the second set, 1000 trajectories for each value of
$b$, in intervals of 0.5 \AA\/, at a given temperature was run for
$t_{\rm fin} = 50$ ps using stratified sampling. All these simulations
were carried out with the ASE simulation
environment,\cite{larsen2017atomic} in the $NVE$ ensemble and with
$\Delta t = 0.1$ fs. The simulations were terminated after $t_{\rm
  fin}$ or when the distance between the oxygen and carbon atom was
larger than 15 \AA\/. The upper limit ($b_{\rm max}$) was determined
from considering the opacity function, see inset Figure
\ref{fig:fig1}. Then, $b$ was drawn between 0 and $b_{\rm
  max}$. All runs that were carried out are summarized in Tables S2 and S3.\\

\subsection{Statistical Modeling}
For the statistical modeling the stationary points along the reaction
coordinate for MgO$^+$ + CH$_4$ were calculated at the TPSS0/TZVP
level using Gaussian 09\cite{g09} and the reported energies have been
zero-point corrected employing harmonic frequencies.  Stable
intermediates and transition states were verified by the calculation
of zero or a single imaginary harmonic frequency, respectively.
Intrinsic reaction coordinate calculations verified the connections
between stationary points.  \\

\noindent
The calculated energies and vibrational and rotational frequencies
along the reaction coordinate were inputs to statistical modeling of
the reaction, described in detail elsewhere.\cite{Sweeny2017} Briefly,
formation of an initial intermediate is determined using capture
theory and the simplified statistical adiabatic channel model
(SSACM),\cite{Stevens2009,Troe2009} with reactant internal and
collision energies varied over thermal distributions in a stochastic
manner.  Intermediates are assumed to be sufficiently long-lived that
the fundamental statistical assumption of energy redistribution is
met, and the fate of the intermediate determined by calculated
unimolecular rate curves as a function of both energy and angular
momentum, ($E, J$), for all exit
channels.\cite{Olzmann1994,Olzmann1992} Importantly, ion-molecule
reactions often involve barrierless formation of entrance complexes,
leading to competition between a ``loose'' dissociation channel and a
``tight'' isomerization transition state; ``loose'' and ``tight''
states are affected differently by angular momentum, and proper
consideration of $J$ is necessary.\cite{TROE198717} The trajectories
are followed until separated products are formed, reactants are
re-formed, or an intermediate stabilized through collision with a
buffer gas.  The process is repeated until sufficient statistics have
been accumulated, which typically requires 10$^4$ to 10$^5$ runs. The
resulting calculated rate coefficients and product branching fractions
are compared with the experimental data.\\

\section{Results and Discussion}
\begin{table}[h]
\small
  \caption{Measured and computed (capture rate from 50 ps simulations)
    rate coefficients and product branching fractions for the reaction
    of MgO$^+$ + CH$_4$ from 300 to 600 K. Literature values are shown
    in brackets.\cite{Schroder2006} $k_{\text{tot}}$ in $\times
    10^{-10}$ cm$^{3}$ s$^{-1}$. Reaction Efficiency is defined as
    ($k_{tot}/k_{LGS}$)$^a$. $^a$Langevin-Gioumousis-Stevenson capture
    rate.\cite{Gioumousis1958}}
\label{tab:tab1}
  \begin{tabular*}{1.0\textwidth}{@{\extracolsep{\fill}}l|ll|l|ll}
\hline
& & & &\multicolumn{2}{l}{Product branching fractions} \\
\hline
$T$(K) & $k_{\text{tot}}$  &   $k_{\text{tot}}^{\rm NN}$ & \text{Efficiency}
  & Mg$^+$ + CH$_3$OH & MgOH$^+$ + CH$_3$ \\
\hline
300 & 5.9 $\pm$ 1.5  & 2.46 & 53\%  & 0.08 $\pm$ 0.03  & 0.92 $\pm$ 0.03  \\
 & \textbf{[3.9 $\pm$ 1.3]}& & \textbf{[$\sim$40\%]} & \textbf{[$<$ 0.02]} &  \textbf{[$>$ 0.98]}\\
 400 & 6.2 $\pm$ 1.6 & 2.37 & 55\% & 0.07 $\pm$ 0.02 & 0.93 $\pm$ 0.02\\
 500 & 4.8 $\pm$ 1.2 &2.34& 43\% & 0.05 $\pm$ 0.03 & 0.95 $\pm$ 0.03\\
 600 & 4.4 $\pm$ 1.1 &1.88& 39\% & 0.05 $\pm$ 0.04 & 0.95 $\pm$ 0.04\\ 
 700 & &1.47&  &   & \\ \hline
    \hline
\end{tabular*}
\end{table}

\begin{figure}[h!]
\centering
\includegraphics[width=0.9\linewidth]{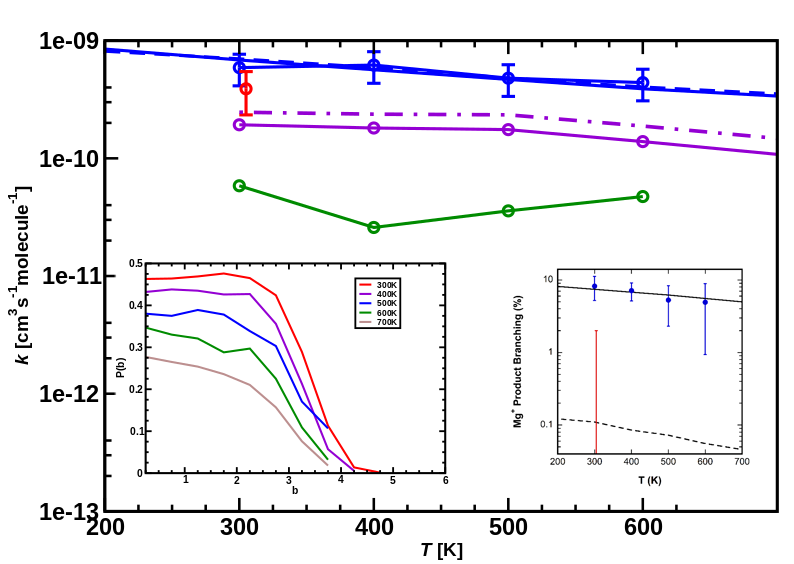}
\caption{Measured rate coefficients (blue circles), and previously
  reported value (red circle)\cite{Schroder2006} for reaction (1),
  MgO$^+$ + CH$_4$, from 300 to 600 K. Best-fit statistical modeling
  assuming TS1 as rate-limiting TS1 at --0.38 eV below the entrance
  channel (solid line) or capture limited (dashed line,
  barrier-independent). The Langevin limit is at $1.1 \times
  10^{-9}$cm$^3$s$^{-1}$molecule$^{-1}$. Calculated rate coefficients
  for temperatures from 300 to 700 K ($\Delta T = 100$ K) MS-ARMD
  (green) and from the NN (solid violet line for Boltzmann rate and
  dashed dotted violet line for the capture rate). The TS from the NN
  lies --0.09 eV (--2.1 kcal/mol) below the entrance channel, see
  black line in Figure \ref{fig:fig8}. The inset reports the opacity
  function calculated from PhysNet at different temperatures.}
\label{fig:fig1}
\end{figure}

\noindent
The measured rate coefficients and product branching fractions for
reaction (1) are listed in Table \ref{tab:tab1} and shown in
Figure \ref{fig:fig1}.  The reaction proceeds at approximately
half of the collision rate at 300 K and decreases $\propto T^{-0.5 \pm
  0.2}$.  The current measurement is somewhat higher than, but not
inconsistent with, the previously reported value.\cite{Schroder2006}
The reaction was found to primarily result in hydrogen atom transfer,
although 8 \% of reactions proceeded by oxygen transfer resulting in
the Mg$^+$ + CH$_3$OH product channel at 300 K.  This is in mild
disagreement with previous reports, which placed an upper limit of 2\%
branching to this channel at 300 K.  The previous work also reports
trace amounts of MgCH$_2^+$ + H$_2$O which was not observed in the
present experiments.\cite{Schroder2006}\\

\subsection{Statistical Modeling}
\begin{figure}[h]
\centering
  \includegraphics[width=\linewidth]{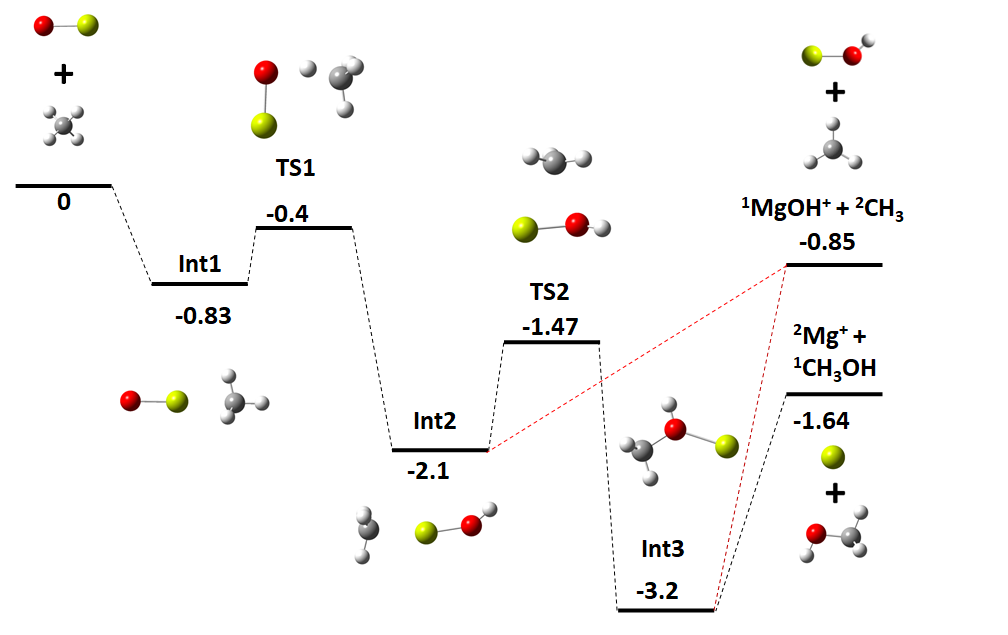}\\
\caption{Energy profile for reaction (1), MgO$^+$ + CH$_4$, calculated
  at the TPSS0/TZVP level; calculated zero-point corrected energies
  (eV) are indicated for each stationary point.}
\label{fig:fig2}
\end{figure}

For the statistical modeling of the main reaction channels the
relevant stationary points were calculated at the TPSS0/TZVP level,
see Figure \ref{fig:fig2}. This profile is qualitatively consistent with
that previously reported calculated at the MP2/6-311+G(2d,2p)
level\cite{Schroder2006} and the results from Figure
\ref{fig:fig8}. The initial encounter complex formed is with the
methane complexed to the metal, bound by 0.83 eV which is 0.43 eV (9.2
kcal/mol) above the van der Waals complex, consistent with the results
from the higher-level calculations. No other bound entrance complexes
were identified.  The reaction proceeds through a four-centered
transition state (TS1) to an insertion complex (INT2) that is
equivalent to a methyl bound to MgOH$^+$.  No TS corresponding to a
direct activation of the C-H bond by end-on attack by the oxygen as
mentioned in previous work,\cite{Schroder2006} was identified;
although the possibility of such a mechanism is discussed below.  From
INT2, competition occurs between direct dissociation into the MgOH$^+$
+ CH$_3$ product, or traversing a transition state corresponding to a
long range methyl migration (TS2) resulting in a deeply bound
Mg-CH$_3$OH$^+$ (INT3) complex.  INT3 can directly dissociate into
either the Mg$^+$ + CH$_3$OH or the MgOH$^+$ + CH$_3$ product
channels, with the former calculated to be preferred by 0.79 eV, just
outside of agreement with experimentally derived reaction energetics
of 0.57 $\pm$ 0.18 eV.\\

\noindent
The experimental kinetics were compared to statistical modeling of the
calculated reaction coordinate.  Treating the MgO$^+$ + CH$_4$
association/dissociation into INT1 using phase-space theory (PST) and
TS1 using Rice-Ramsperger-Kassel-Marcus (RRKM) theory models the total
rate coefficients well (Figure \ref{fig:fig1}, solid blue line) by
minimally adjusting the energy of TS1 to $-0.38 \pm 0.10$ eV relative
to reactants. The calculated lifetime of INT1 is $>> 10^{-10}$ s at
thermal energies, long enough that intra-molecular vibrational energy
redistribution should likely be complete and the fundamental
assumption of statistical theory valid.  An alternative direct
hydrogen abstraction channel was previously calculated to involve a TS
0.1 eV above TS1.  That statistical behavior is expected based on the
calculated INT1 lifetime along with the agreement between the
experiment and the model suggests that the reaction does proceed
statistically, primarily through TS1 at these energies. Shaun: Still,
it is possible that a smaller fraction proceeds by direct hydrogen
abstraction although this channel is not included in the present model
and the agreement between model and experiment without this channel
supports its negligible impact on the results.\\

\noindent
MgO$^+$ + CH$_4$ association/dissociation may proceed with rate
coefficients below PST due to ``rigidity'' which is related to angular
anisotropy of the potential along the dissociation
coordinate.\cite{Troe2009} This is treated as an adjustable parameter
in the following. The SSACM formalism is used to calculate
unimolecular rate curves for loose dissociations from the PST-limit,
where transitional modes are treated as product rotations, down to the
RRKM-limit, where transitional modes are treated as fully conserved
vibrations, as a function of a single ``rigidity factor''.  Increasing
rigidity in the entrance channel necessarily lowers the modeled
efficiency of the reaction as well as imparting a small negative
temperature dependence due to the magnitude of the effect of rigidity
increasing with energy. While lowering the energy of TS1 below --0.38
eV increases the reaction efficiency above the experiment, this
increase may be compensated for by applying a rigidity factor
(lowering the capture rate below the PST value) while maintaining a
temperature dependence consistent with experiment. As a result, in the
present case the modeling provides only an upper limit ($-0.38 \pm
0.1$ eV) for the TS1 energy. In light of this, scans of the angular
and dissociation coordinate were undertaken (Figure
\ref{fig:fig4}). These results are not consistent with the large
amount of anisotropy required to render this reaction capture
controlled. Furthermore, previous work on very similar associations
(FeO$^+$, NiO$^+$, and MnO$^+$ with CH$_4$) were all well modelled at
the PST limit suggesting minimal, if any, impact of
rigidity.\cite{Ard2014,McDonald2018,Sweeny2019} \\

\begin{figure}[h]
\centering
  \includegraphics[width=\linewidth]{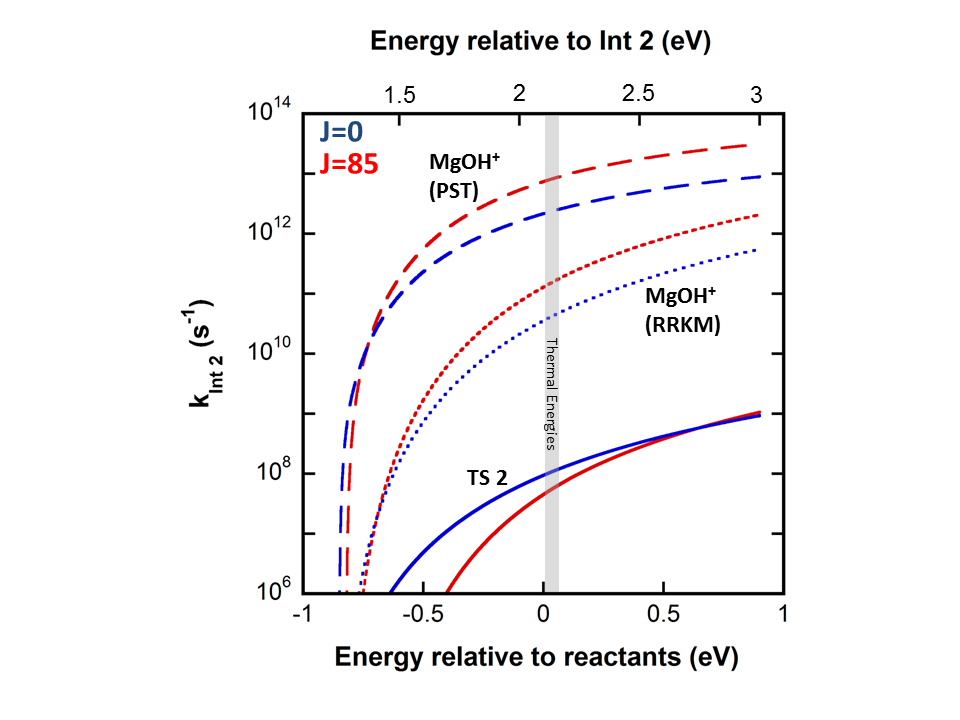}\\ 
\caption{Statistical Rate curves calculated using SSACM for $J = 0$
  (blue) and $J = 85$ (red) to exiting INT2 by TS2 (solid lines),
  dissociation into the MgOH$^+$ product channel at the PST-limit
  (dashed lines), and dissociation into the MgOH$^+$ product channel
  at the RRKM-limit (dotted lines).  The grey shadowed region
  indicates the energetic region of importance at 300 K. }
\label{fig:fig3}
\end{figure}

\begin{figure}[h!]
\centering
\includegraphics[width=0.7\linewidth]{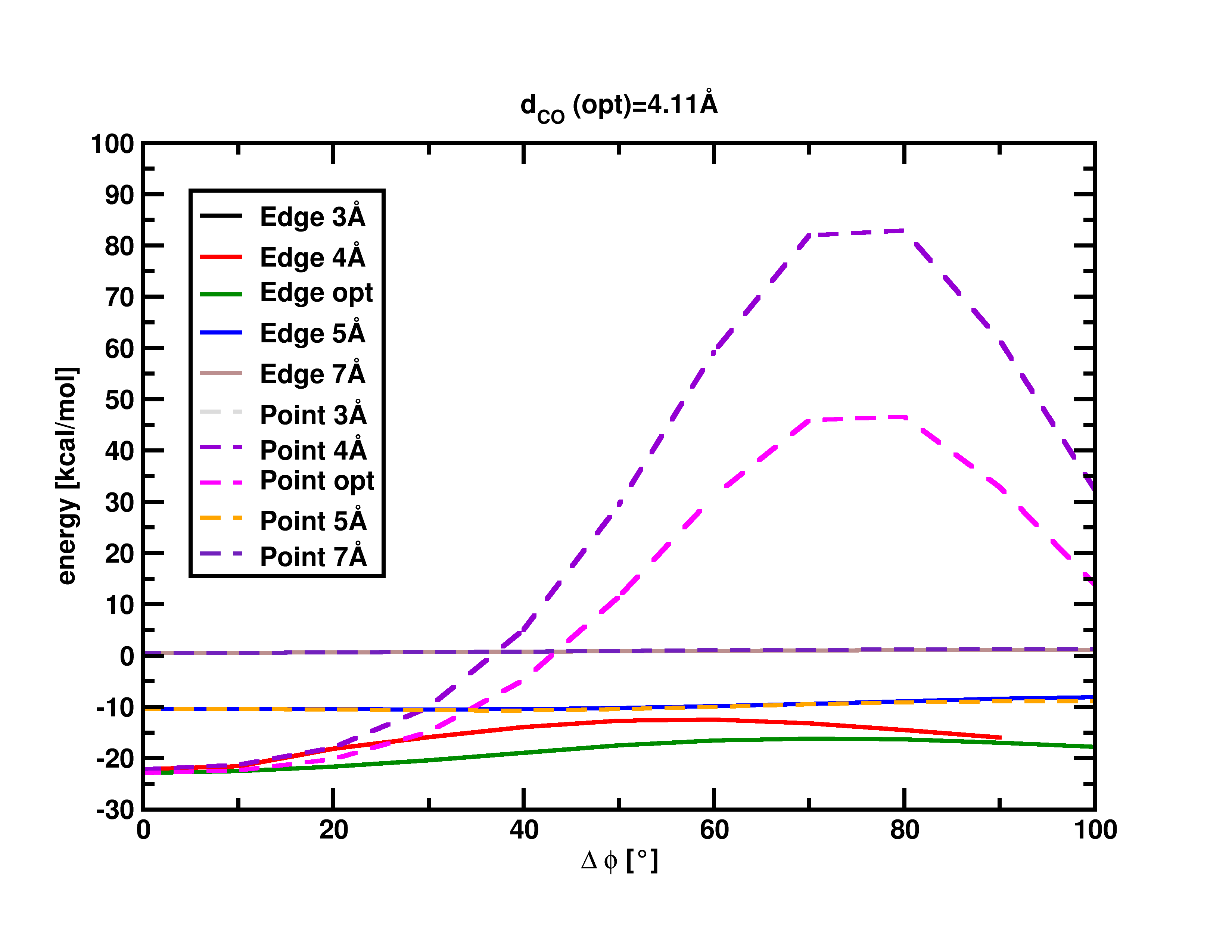}\\
\caption{MP2/6-311+G(2d,2p) scans along the angular and dissociation
  coordinate (defined as the CO distance). ``Edge'' refers to the
  rotation through an imaginary line between two CH$_4$ hydrogen
  atoms, ``points'' refers to a rotation around one of the CH$_4$
  hydrogens.}
\label{fig:fig4}
\end{figure}

\noindent
The product branching between reactions Eq. \ref{eq:1} (leading to
MgOH$^+$) and Eq. \ref{eq:2} (leading to Mg$^+$) is determined by
competition after the rate-limiting transition state (TS1).  Figure
\ref{fig:fig3} displays several calculated rate curves for exiting
INT2 at both $J = 0$ and $J = 85$. The latter is near the peak of the
angular momentum distribution for this reaction at 300 K. For
simplicity the rate curves shown are at the same energetic threshold,
--0.85 eV relative to reactants. Three sets of curves are shown, a)
assuming a loose dissociation (bond fissure) by phase space theory, b)
dissociation at the RRKM limit (extremely tight), and c) traversing
the TS2 barrier. The PST curves (dashed) are orders of magnitude
faster ($\sim$10$^4$) than those for traversing TS2 (solid). This is
independent of any adjustment to the various energetics. Such
adjustments can be visualized by translating the curves left or right
relative to each other. This modeling predicts essentially unit
branching to MgOH$^+$, inconsistent with the several percent of Mg$^+$
observed.  Increasing the ``rigidity'' of the dissociation bends the
dissociation rate curve between the PST and RRKM limits shown in
Figure \ref{fig:fig3}.  Employing ``rigidity'' reduces the
entropic preference of this channel over traversing TS2, see Figure
\ref{fig:fig3}. However even at the RRKM-limit, less than 1\% of
products are predicted to form Mg$^+$ without a significant energetic
adjustment. In fact energy adjustments only are not sufficient to
model the product branching observed. For this one must treat not only
dissociation to the MgOH$^+$ + CH$_3$ product from INT2 at the RRKM
limit, but from INT3 as well. Then the majority of complexes that make
it to INT3 result in Mg$^+$ + CH$_3$OH formation.\\

\noindent
Figure \ref{fig:fig1} (inset) shows the observed product branching
as well as modeled predictions employing the RRKM limit for the
dissociation to MgOH$^+$ from both INT2 and INT3 while keeping the two
channels isoenergetic at -0.85 eV relative to reactants (black dashed
line).  The product branching was only modeled well (inset Figure
\ref{fig:fig1}, black solid line) when further lowering the energy
of TS2 to --1.55 eV, while the energies of both product channels are
left as calculated.  While this produces a suitable fit to the data,
it remains unsatisfactory.  Even treating dissociations to MgOH$^+$ +
CH$_3$ from both INT2 and INT3 at the extreme of the RRKM limit, a
good fit to the data requires that the energy of TS2 be no higher than
--1.55 eV relative to reactants, which would suggest that TS2 is
heavily stabilized (by at least 0.7 eV) relative to the MgOH$^+$ +
CH$_3$ channel.  This seems unlikely given that TS2 corresponds to a
long range migration of the methyl group from the metal to the
hydroxyl site, with minimal deformation to either the methyl or metal
hydroxide unit, and would therefore be expected to be much closer to
the energy of the MgOH$^+$ + CH$_3$ channel.  While current DFT
calculations have this TS lying 0.62 eV below this product channel,
previous calculations using MP2 show them much closer to isoenergetic,
with a difference of only 0.01 eV.  Additionally, increasing the size
of the basis set from TZVP to def2-TZVP or better accounting for long
range interaction by using the CAM-B3LYP method raises the calculated
energy of TS2 significantly. The parameters required to model the
product branching under statistical assumptions appear possibly
unphysical. This suggests that after passing TS1 the reaction may in
fact proceed non-statistically.\\

\begin{figure}[h]
\centering
  \includegraphics[width=\linewidth]{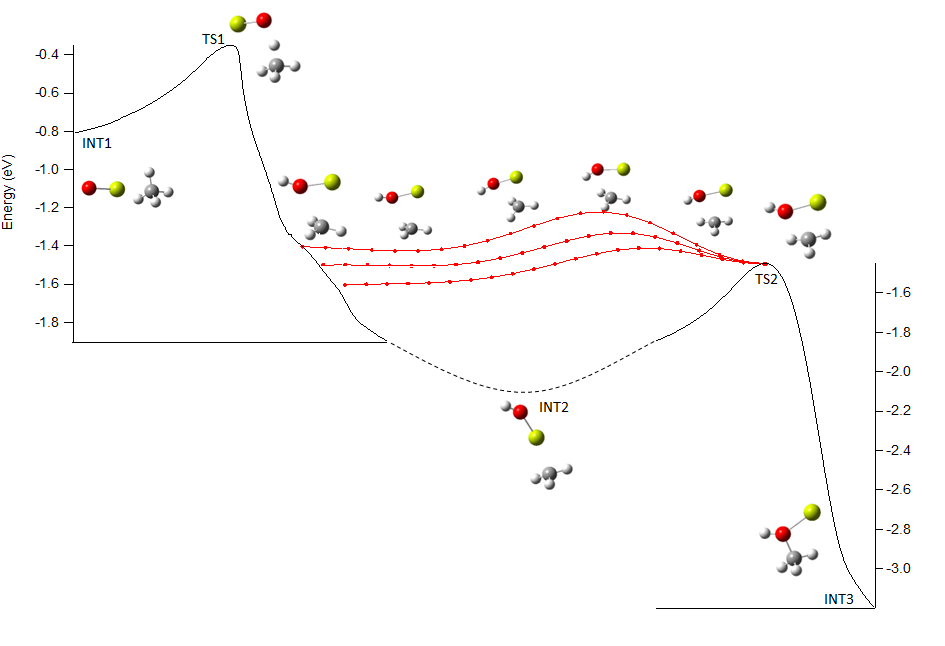}\\ 
\caption{Calculated IRCs from TS1 and TS2 (solid lines; dashed line is
  interpolation to INT2) and stationary point structures are shown.
  Structures at the shoulder along the IRC and along paths connecting
  to TS2 are shown to depict possible reaction path bifurcation
  directly connecting TS1 and TS2 (see text). }
\label{fig:fig5}
\end{figure}

\noindent
Therefore, the potential surface in the post-transition state region
was explored further. A shoulder along the IRC from TS1 to INT2 was
observed with a structure very similar to that of TS2, see Figure
\ref{fig:fig5}. The PES between this shoulder and TS2 was explored by
incrementally adjusting a set of internal coordinates from several
geometries along the IRC to those for TS2 (Figure \ref{fig:fig5}, red
points). The PES in this region is rather flat providing an
energetically feasible route directly to TS2 and thus INT3,
effectively bypassing INT2, see Figure \ref{fig:fig5}. This feature
could be a reaction path bifurcation (RPB), an example of
post-transition state non-statistical dynamics. An RPB can be
visualized as a ridge in the potential running from one TS to another,
separating two valleys containing the intermediates.\\

\noindent
The key kinetic step in determining the branching becomes the
competition between following the minimum energy pathway to INT2,
which would produce entirely the MgOH$^+$ product, or traversing
directly to TS2 and then INT3, where the competition between the two
product channels is more robust. The presence of an RPB provides an
explanation for why statistical assumptions are valid for reaction (1)
prior to the rate-limiting TS1, but fail after that point. Calculation
of the full potential surface in the region of TS1, INT2, TS2 coupled
with quasiclassical trajectory calculations or exploration by direct
dynamics calculations can verify the presence of an RPB for this
system and provide a fuller exploration of the interesting
non-statistical dynamics occurring.  \\

\subsection{Reactive MD Simulations}
\begin{figure}[h!]
\centering \includegraphics[width=\linewidth]{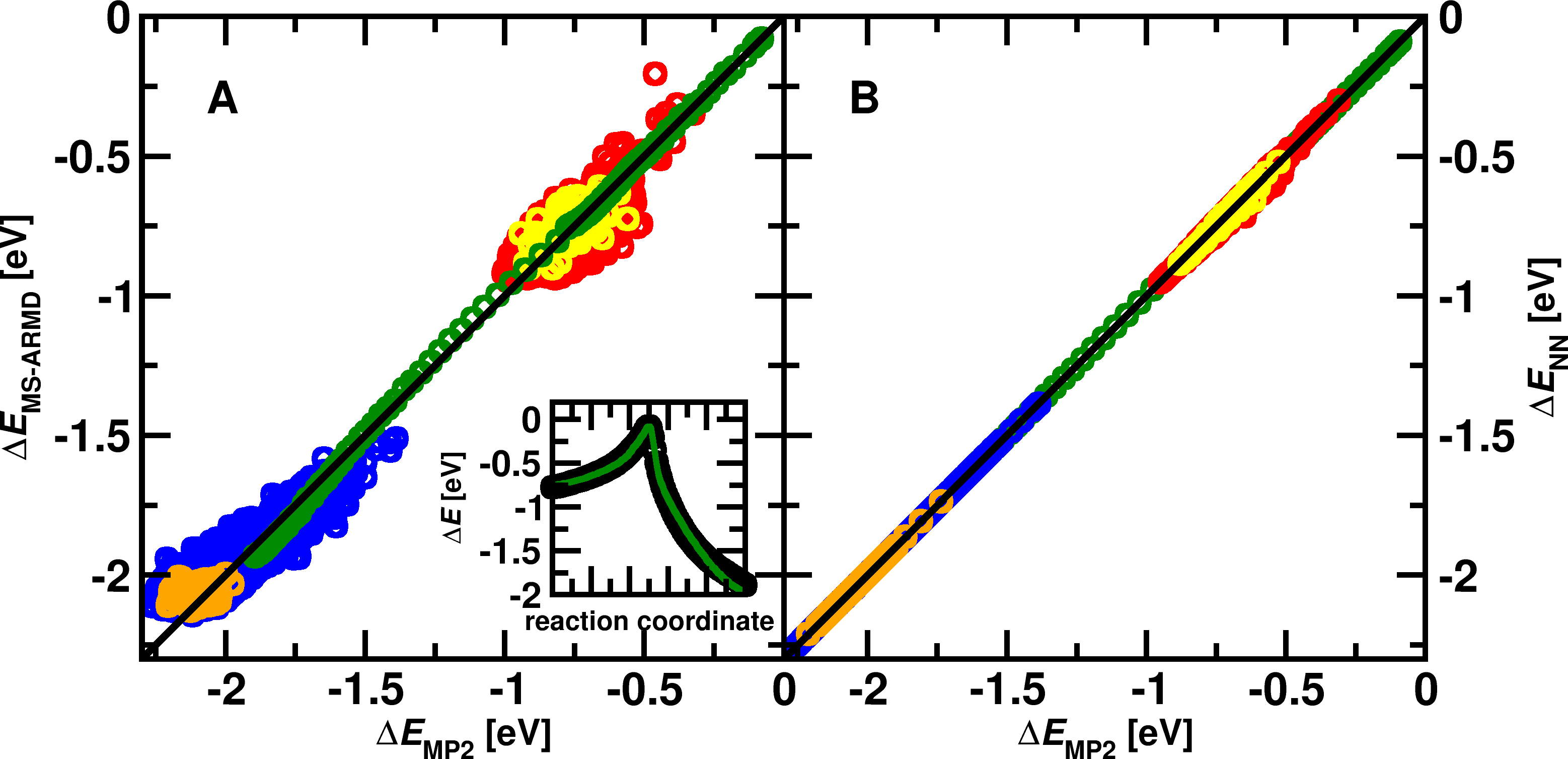}
\caption{Panel A: Correlation between 1505 \textit{ab initio}
  reference energies used in the parametrization and the fitted
  MS-ARMD force field of the reactant complex (red, RMSD = 0.04 eV;
  0.89 kcal/mol) and of 85 structures for validation (yellow, RMSD =
  0.07 eV; 1.55 kcal/mol). Also, the correlation for 1730 \textit{ab
    initio} energies for the product complex (blue, RMSD = 0.05 eV;
  1.10 kcal/mol) and of 91 structures for validation (orange, RMSD =
  0.06 eV; 1.48 kcal/mol) are reported. Green dots show the reference
  IRC compared with that from the fitted MS-ARMD force field (RMSD =
  0.02 eV; 0.49 kcal/mol). The insets shows the reference IRC (black)
  and the MS-ARMD energies (green). Panel B: Correlation of energies
  calculated at the MP2/aug-cc-pVTZ level of theory and predictions of
  the NN for the same sets of structures (using the same color code).
  Note that these structures were not used during the training of the
  NN. (red, RMSD = 0.0124 eV; 0.285 kcal/mol) (yellow, RMSD = 0.0115
  eV; 0.265 kcal/mol) (blue, RMSD = 0.000157 eV; 0.00363 kcal/mol)
  (orange, RMSD = 0.000200 eV, 0.00462 kcal/mol) (green, RMSD =
  0.00337 eV; 0.0777 kcal/mol).}
\label{fig:fig6}
\end{figure}

The quality of the two reactive PESs is reported in Figure
\ref{fig:fig6} with a typical ${\rm RMSD} \sim 1$ kcal/mol for the
MS-ARMD PES and a much lower ${\rm RMSD} \sim 10^{-2}$ kcal/mol for
the NN-trained PES. The IRC between Int+ and TS1 (inset in Figure
\ref{fig:fig6}A) is well reproduced by the MS-ARMD PES. Note that the
structures shown in Figure \ref{fig:fig6}B were not used for training
the NN and act as an additional test set. The quality of the NN-fitted
PES is evidently superior to that from the MS-ARMD fit in particular
because globally, it correctly describes the relative energetics of
the entrance channel, the van der Waals minimum (INT1) and TS. On the
other hand, the MS-ARMD PES is computationally considerably more
efficient.\\

\noindent
For the direct process MgO$^+$ + CH$_4 \rightarrow$ MgOH$^+$ + CH$_3$,
reactive MD simulations have been carried out using the parametrized
MS-ARMD PES. The rate coefficients for the investigated reaction for
300 to 600 K are shown in Figure \ref{fig:fig1}. Compared to
experimental rates determined from electrospray ionization the results
from MS-ARMD (5.9 $\times 10^{-11}$ cm$^{-3}$ s$^{-1}$ at 300 K) are
about one order of magnitude smaller than previously experimentally
determined $k$ values ($3.9 \pm 1.3 \times 10^{-10}$ cm$^{-3}$
s$^{-1}$ at 300 K),\cite{Schroder2006} as well as the measured rates
in the present work ($5.9 \times 10^{-10}$ cm$^{-3}$ s$^{-1}$ at 300
K; see Figure \ref{fig:fig1}). MS-ARMD simulations yield a mild
$T-$dependence, in qualitative agreement with experiment.\\

\begin{figure}[h]
\centering \includegraphics[width=\linewidth]{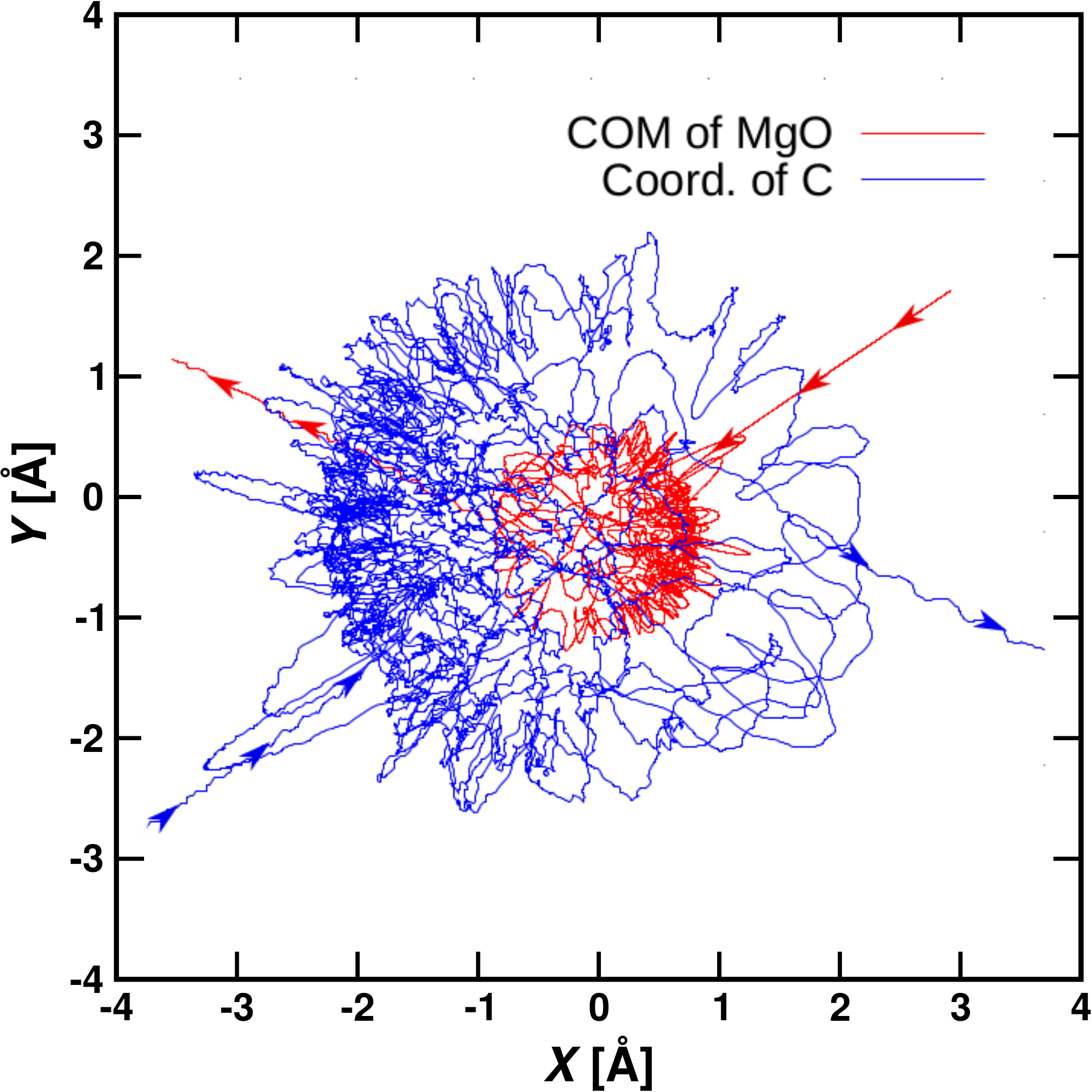}\\
\caption{Reactive trajectory calculated from the NN-PES at 500 K with
  $b = 0.5$ a$_0$. The two reacting species ``roam'' about one another
  before the reaction occurs. The incoming and outgoing parts of the
  trajectory are indicated by arrows. The trajectories follow the
  center of mass of MgO (red) and the carbon of CH$_4$ (blue).}
\label{fig:fig7}
\end{figure}

\noindent
Using the NN-trained PES the temperature-weighted rate is about a
factor of 5 smaller than that from experiment and the $T-$dependence
follows the experimental rate. A typical reactive trajectory from such
simulations is shown in Figure \ref{fig:fig7}. The NN also includes
the possibility for dissociation to Mg$^+$ + CH$_3$OH which was indeed
found for $\sim 10$ out of 40,000 trajectories. This is consistent,
but not in quantitative agreement, with a low-probability channel
found in the experiments, see Table \ref{tab:tab1}. Considerably
more statistics would be required for a direct comparison.\\

\noindent
Despite the different quality and shortcomings of the two PESs,
simulations with both approaches (MS-ARMD and NN) underestimate the
experiments and requires an explanation. Two possibilities are
explored in the following: first the accuracy of the electronic
structure calculations is considered and secondly the fact that a
large fraction of the NN simulations is trapped in INT1 (3545, 700,
375, and 507 for $T = 300, 400, 500$, and 600 K, respectively) but
could potentially react on longer time scales and thus contribute to
the rate.\\

\begin{figure}[h]
\centering
  \includegraphics[width=\linewidth]{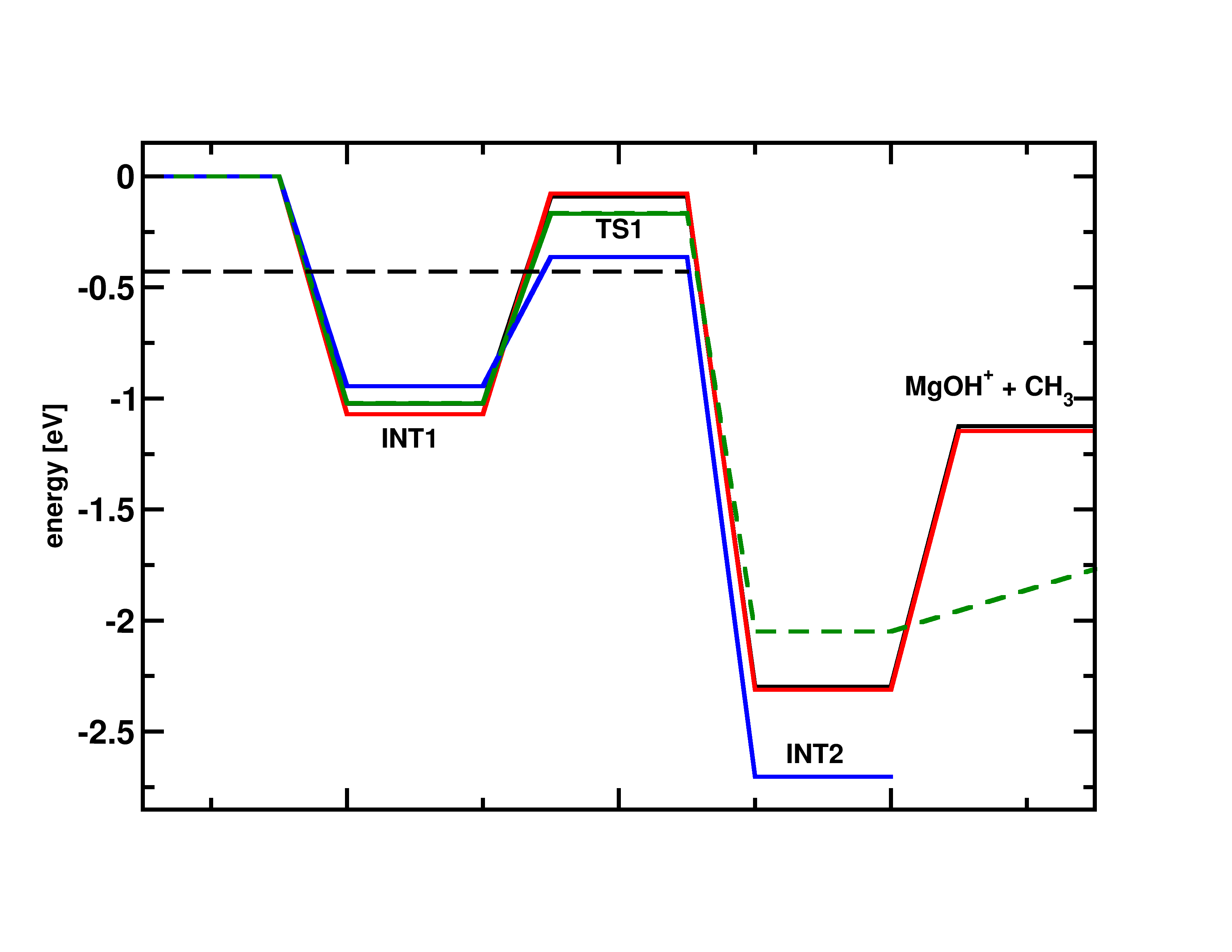}\\ 
\caption{Potential energy surface (in eV) for the reaction of MgO$^+$
  with CH$_4$ calculated at the MP2/6-311+G(2d,2p) (red) and
  MP2/aug-cc-pVTZ (black) levels of theory. Furthermore the results
  from CCSD(T)/aug-cc-pVTZ (green; solid from Gaussian09\cite{g09} and
  dashed from Molpro\cite{molpro}) and CASSCF/VDZ calculations
  (violet). The barrier height from statistical modeling of the
  experimental rates is the black dashed line. From INT2 the reaction
  can progress via a second TS and a stable intermediate, see Figure
  \ref{fig:fig5}.}
\label{fig:fig8}
\end{figure}

\noindent
A difference of one order of magnitude in the rate corresponds to a
change of $\sim 0.07$ eV ($\sim 1.5$ kcal/mol) in the relevant barrier
assuming a single pathway and the validity of transition state
theory. In other words, as all computed rates are smaller by that
amount, the relevant barrier between INT1 and TS1 (see Figures
\ref{fig:fig2} and \ref{fig:fig6}) for the reaction is probably
overestimated from the electronic structure calculations. In order to
better understand this, the energies for the stationary points were
also determined at the CCSD(T)/aug-cc-pVTZ level of theory, see green
trace in Figure \ref{fig:fig8} which yields a barrier between INT1
and TS1 of 0.86 eV (19.8 kcal/mol reference energy to reactants) which
is smaller than that at the MP2 level, but still overestimates the
barrier height from analyzing the experimental rates. \\

\noindent
Typically, calculations at the CCSD(T) level of theory as employed
here should provide accuracies better than 0.05 eV (1
kcal/mol).\cite{Parthiban2001,Claeyssens2006} Hence a difference of up
to 0.22 eV (5 kcal/mol) between the barrier height from analyzing
experiment (--0.38 eV) and the computed one (ranging from MP2 = --0.07
eV to CCSD(T) = --0.17 eV) points towards another source of
error. Hence, the T1 diagnostic as an indication for single- versus
multi-reference character was considered.\cite{Lee1989} At the
CCSD/aug-cc-pVTZ level of theory T1 of TS1 is 0.03 which exceeds the
recommended value of 0.02 for single-reference
character.\cite{Lee1989} For the reactant and product states the T1
diagnostic is considerably smaller than 0.02 and thus, a
single-reference calculation should provide good results. At the
CASSCF level of theory the TS structure is about 0.35 eV (8 kcal/mol)
below the entrance channel. This is consistent with the statistical
modeling from the present work. Hence, it is concluded that including
the multi-reference character of the wavefunction is likely to lower
the barrier and to increase the rate, in agreement with the findings
in Figure \ref{fig:fig1}.\\

\noindent
Next, for the reactive trajectories using the NN PES, the influence of
the trajectories still trapped after 50 ps and 500 ps, respectively,
in the van der Waals well on the total rate was quantified. Because
the NN-PES supports the van der Waals, pre-reactive complex, an
appreciable fraction of simulations still remain trapped there. The
trapped fraction ranges from 90 \% at 300 K to 40 \% at 700 K, see
Table S2. As the forward barrier (towards TS1) is lower than
the reverse barrier (back to separated reactants) it is conceivable
that an appreciable fraction of these trajectories will still react
towards products, but on considerably longer time scales, due to
IVR. Following previous work,\cite{Koner2014} the (unconverged) rate
determined from NN-MD simulations was corrected by multiplying with a
Boltzmann factor $\frac{p_i}{p_j} = e^{\frac{\varepsilon_j -
    \varepsilon_i}{k_B T}}$. Here, $p_i/p_j$ is the ratio between the
product and total probabilities, $\varepsilon_j$ and $\varepsilon_i$
are the energies of the entrance channel and the TS energy (1.02 eV
and 0.93 eV), and $k_B$ and $T$ the Boltzmann constant and the
temperature, respectively. Correcting the rate determined from direct
sampling yields rates on the order of $\approx 2.3 \times 10^{-10}$,
see Figure \ref{fig:fig1} (solid violet for Boltzmann rate and
solid dash-dotted for the capture rate), which agrees more favorably
with the experimental rates, especially for low temperatures. The
correction can also be considered as a way to extrapolate the rate
from sampling a finite time interval to infinitely long
simulations. Still, these rates are subject to the overestimation of
the barrier between Int1 and TS1 due to the multireference character
of TS1.\\

\section{Conclusion}
In this work, the temperature dependence of the rate coefficient and
the product branching ratio for the reaction of MgO$^+$ + CH$_4$ have
been determined from experiments and analyzed with reactive MD
simulations (using MS-ARMD and a NN-trained PES) together with
statistical modeling. The primary rate limiting step at thermal
energies was confirmed to be a four-centered metal mediated C-H
activation as had previously been postulated.\cite{Schroder2006} With
increasing temperature the thermal rate decreases slightly and
formation of Mg$^+$ reduces from 8 \% at 300 K to 5 \% at 700 K.\\

\noindent
The statistical modeling of the thermal rate yields a submerged
barrier, --0.38 eV below the entrance channel. The product branching
was only reproduced within the limits of the modeling with a
significant stabilization of a TS corresponding to a long range methyl
migration, as well as substantial ``rigidity'' imparted to
dissociation into the MgOH$^+$ + CH$_3$ channel.  Direct C-H
activation by end on attack of the O atom\cite{Schroder2006} was not
found in the current calculations and the good agreement with
experimental rates without this channel suggests minimal
impact. Finally, calculations suggest the possibility of a post TS
bifurcation in the potential surface which could be responsible for
the observed product branching.\\

\noindent
Thermal rates were also determined from atomistic simulations using
two reactive force fields, one based on MS-ARMD and the other one
using a NN trained to extensive reference data. The MS-ARMD
simulations are run with excess collisional energy because the global
PES can not describe the relative energetics of all states
involved. The rates determined from MS-ARMD show a negative
$T-$dependence, consistent with experiment, but are about one order of
magnitude smaller. Rates from the simulations using the NN also show
the correct $T-$dependence and are smaller by about a factor of 5
compared with experiment. Both PESs (MS-ARMD and NN) overestimate the
height of the submerged barrier which was traced back to the
electronic structure calculations. Examination of the energetics
suggests that the height of TS1 is affected by multi-reference
effects. The MD simulations using the NN-trained PES also find
formation of Mg$^+$ but the fraction of these trajectories is much
smaller than that observed experimentally.\\

\noindent
Overall, by combining experiment and computational modeling a
comprehensive understanding of the key step of thermal activation of
methane by MgO$^+$ was obtained. The rate limiting step involves a
submerged barrier associated with a four-centered transition state as
has been previously stipulated by calculations\cite{Schroder2006}
which leads to a negative temperature dependence of the rate. This is
supported by the statistical modeling, the electronic structure
calculations, and the reactive molecular dynamics simulations. The
kinetics controlling the competition between energetically available
product channels is poorly reproduced by statistical methods, possibly
due to a bifurcation in the potential surface after the rate limiting
step (TS1) leading to non-statistical behavior.\\

\section*{Conflicts of interest}
There are no conflicts of interest to declare.

\section*{Acknowledgments}
This work was supported by the Swiss National Science Foundation
through grants 200021-117810, 200020-188724, and the NCCR MUST (to
M.M.) and by the Air Force Office of Scientific Research
(AFOSR-19RVCOR042, to S. G. A.). B.C.S. and J.C.S. are supported by
the National Research Council Research Associateship
Program. S.G.A. received support in part through the Institute for
Scientific Research of Boston College under Contract
No. FA9453-10-C-0206.

\bibliographystyle{ieeetr}
\bibliography{refs}

\end{document}

\title{Thermal Activation of Methane by MgO$^+$: Temperature Dependent
  Kinetics, Reactive Molecular Dynamics Simulations and Statistical
  Modeling}

\author{Brendan C. Sweeny} \affiliation{NRC postdoc at Air Force
  Research Laboratory, Space Vehicles Directorate, Kirtland Air Force
  Base, New Mexico 87117}

\author{Hanqing Pan} \affiliation{USRA Space Scholar at Air Force
  Research Laboratory, Space Vehicles Directorate, Kirtland Air Force
  Base, New Mexico 87117}

\author{Asmaa Kassem} \affiliation{USRA Space Scholar at Air Force
  Research Laboratory, Space Vehicles Directorate, Kirtland Air Force
  Base, New Mexico 87117}

\author{Jordan C Sawyer} \affiliation{NRC postdoc at Air Force
  Research Laboratory, Space Vehicles Directorate, Kirtland Air Force
  Base, New Mexico 87117}

\author{Shaun G. Ard} \email[]{rvborgmailbox@us.af.mil}
\affiliation{Air Force Research Laboratory, Space Vehicles
  Directorate, Kirtland Air Force Base, New Mexico 87117}

\author{Nicholas S. Shuman} \affiliation{Air Force Research
  Laboratory, Space Vehicles Directorate, Kirtland Air Force Base, New
  Mexico 87117}

\author{Albert A. Viggiano} \affiliation{Air Force Research
  Laboratory, Space Vehicles Directorate, Kirtland Air Force Base, New
  Mexico 87117}

\author{Sebastian Brickel} \affiliation{Department of Chemistry,
  University of Basel, Klingelbergstrasse 80, CH-4056 Basel,
  Switzerland\\ Present Address: Department of Chemistry - BMC,
Uppsala University, BMC Box 576, 751 23 Uppsala,Sweden}

\author{Oliver T. Unke} \affiliation{Department of Chemistry,
  University of Basel, Klingelbergstrasse 80, CH-4056 Basel,
  Switzerland\\ Present Address: Machine Learning Group, TU Berlin,
  Marchstr. 23, 10587 Berlin, Germany}

\author{Meenu Upadhyay} \affiliation{Department of Chemistry,
  University of Basel, Klingelbergstrasse 80, CH-4056 Basel,
  Switzerland}

\author{Markus Meuwly} \email[]{m.meuwly@unibas.ch}
\affiliation{Department of Chemistry, University of Basel,
  Klingelbergstrasse 80, CH-4056 Basel, Switzerland}

\date{\today}

\maketitle

\section{Quality of the NN-learned force field}

\begin{figure}[h]
\centering
\includegraphics[width=\linewidth]{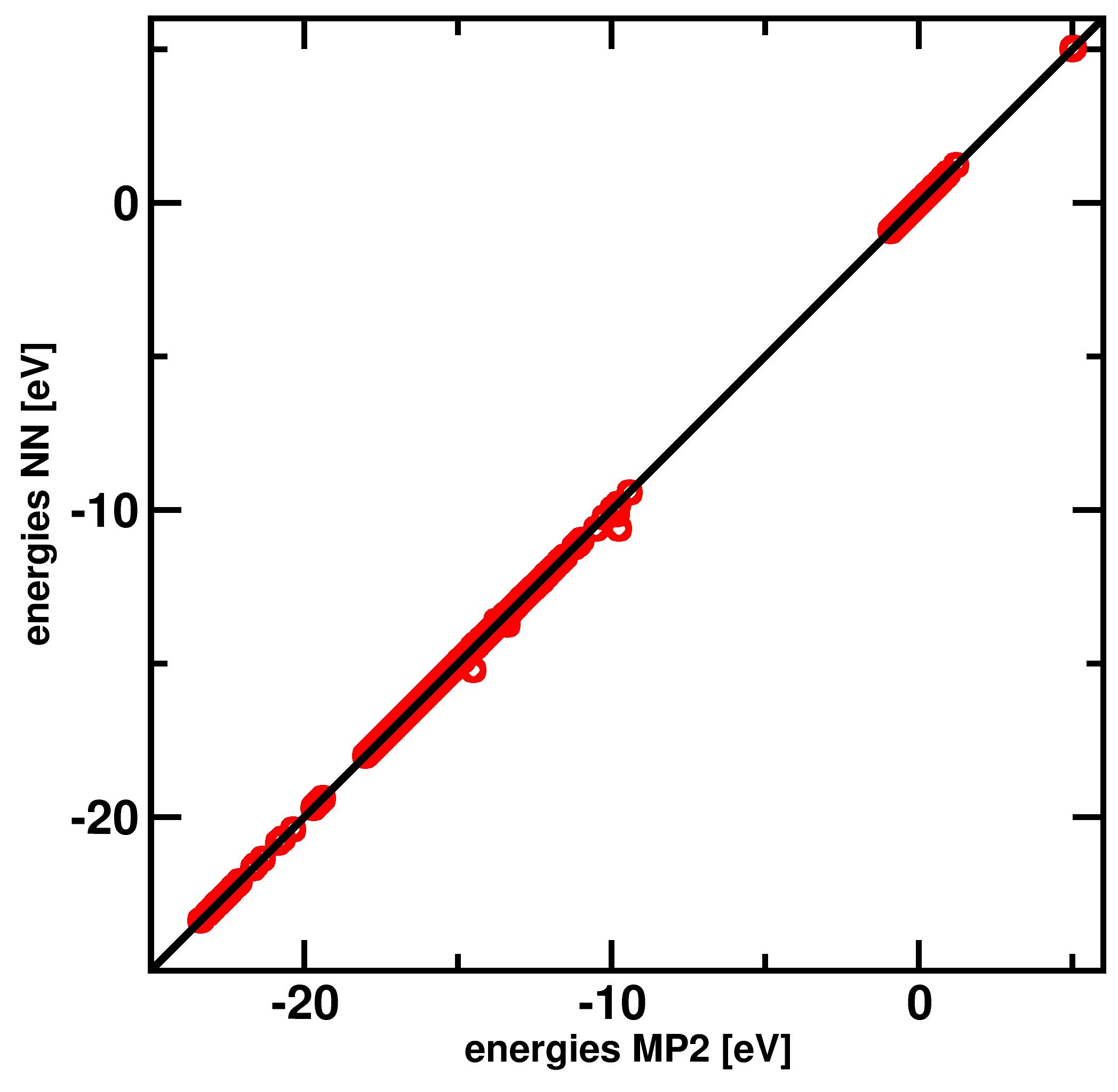}
\caption{Correlation between the energies for 4368 \textit{ab initio}
  reference structures from the test set and the trained neural
  network. Including the separated molecules (MgO$^+$, CH$_4$,
  MgOH$^+$, CH$_3$, CH$_3$OH, Mg$^+$), transition states and complexes
  with $R^2 = 1-4.1*10^{-6}$.}
\label{figsi1}
\end{figure}

\begin{scriptsize}
  
\begin{table}[h!]
  \begin{center}
\caption{List of trajectories, 50 ps in length, from stratified
  sampling of the NN-PES.}
    \begin{tabular}{l|m|j|r|n|q|x}
      \textbf{} & \textbf{$T$[K]} & $b$ & \textbf{Reactive} & \textbf{Complex} & \textbf{Nonreactive}& \textbf{Total}\\
       \hline
      1 & 300 &0.0--0.5 & 3 & 1394 & 180 & 1577\\
      2 &     & 0.5--1.0 & 1 & 1492 & 188 & 1681\\
      3 &     & 1.0--1.5 & 1 & 1435 & 164& 1600\\
      4 &     & 1.5--2.0 & 0 & 1297 & 127& 1424\\
      5 &     & 2.0--2.5 & 0 & 1122 & 143& 1265\\
      6 &     & 2.5--3.0 & 0 &  888 & 206 & 1094\\
      7 &     & 3.0--3.5 & 0 &  521 & 420&  941\\
      8 &     & 3.5--4.0 & 0 &  172 &623 &  795\\
      9 &     & 4.0--4.5 & 0 &   17 & 631&  648\\
     10 &     & 4.5--5.0 & 0 &    1 & 476&  477\\
     11 &     & 5.0--5.5 & 0 &    0 & 294&  294\\
     12 &     & 5.5--6.0 & 0 &    0 & 167&  167\\
     \hline
     13 & 400 & 0.0--0.5 & 11 & 1291 & 195& 1497\\
     14 &     & 0.5--1.0 & 9 & 1326 & 268& 1603\\
     15 &     & 1.0--1.5 & 8 & 1152 & 242& 1402\\
     16 &     & 1.5--2.0 & 2 &  982 & 225& 1209\\
     17 &     & 2.0--2.5 & 1 &  812 & 182&  995\\
     18 &     & 2.5--3.0 & 0 &  560 & 261&  821\\
     19 &     & 3.0--3.5 & 0 &  261 & 379&  640\\
     20 &     & 3.5--4.0 & 0 &   48 & 387&  435\\
     21 &     & 4.0--4.5 & 0 &    2 & 191&  193\\
     \hline
     22 & 500 & 0.0--0.5 & 25 & 1054 & 430& 1509\\
     23 &     & 0.5--1.0 & 19 & 1121 & 469& 1609\\
     24 &     & 1.0--1.5 & 13 &  993 & 359& 1365\\
     25 &     & 1.5--2.0 & 5  &   792& 311& 1108\\
     26 &     & 2.0--2.5 & 3  &   582& 322&  907\\
     27 &     & 2.5--3.0 & 0  &   406& 294&  700\\
     28 &     & 3.0--3.5 & 1  &   145& 259&  405\\
     29 &     & 3.5--4.0 & 0  &    66& 257&  323\\
     \hline
     \label{tab:tabs1}
    \end{tabular}
  \end{center}
\end{table}

\begin{table}[h!]
  \begin{center}
     \begin{tabular}{l|m|j|r|n|q|x}
      \textbf{} & \textbf{$T$[K]} & $b$ & \textbf{Reactive} & \textbf{Complex} & \textbf{Nonreactive}& \textbf{Total}\\
       \hline
     30 & 600 & 0.0--0.5 & 53 & 891 & 551 & 1495\\
     31 &     & 0.5--1.0 & 38 & 991 & 528 & 1557\\
     32 &     & 1.0--1.5 & 16 & 811 & 542 & 1369\\
     33 &     & 1.5--2.0 & 8  & 622 & 525 & 1155\\
     34 &     & 2.0--2.5 & 6  & 515 & 405 &  926\\
     35 &     & 2.5--3.0 & 1  & 306 & 407 &  714\\
     36 &     & 3.0--3.5 & 1  & 92  & 355 &  448\\
     37 &     & 3.5--4.0 & 0  & 16  & 238 &  254\\
     \hline
     38 & 700 & 0.0--0.5 & 64 & 697& 782& 1543\\
     39 &     & 0.5--1.0 & 58 & 685& 823& 1566\\
     40 &     & 1.0--1.5 & 27 & 621& 733& 1381\\
     41 &     & 1.5--2.0 & 16 & 484& 639& 1139\\
     42 &     & 2.0--2.5 & 7  & 354& 552&  913\\
     43 &     & 2.5--3.0 & 4  & 209& 509&  722\\
     44 &     & 3.0--3.5 & 1  &  71& 430&  502\\
     45 &     & 3.5--4.0 & 0  &   7& 195&  202\\
     \hline
     \label{tab:tabs1}
    \end{tabular}
  \end{center}
\end{table}

\newpage

\begin{table}[h!]
  \begin{center}
\caption{List of trajectories, 500 ps in length, from uniform
  sampling of the NN-PES.}
    \begin{tabular}{l|m|j|r|n|q|x}
      \textbf{} & \textbf{$T$[K]} & $b$ & \textbf{Reactive} & \textbf{Complex} & \textbf{Nonreactive}& \textbf{Total}\\
       \hline
      1 & 300 &0.0--0.5  &82 & 1140 & 73 & 1295\\
      2 &     & 0.5--1.0 &82 &  680 & 24 & 786\\
      3 &     & 1.0--1.5 & 9 &  474 & 55 & 538\\
      4 &     & 1.5--2.0 & 6 &  342 & 38 & 386\\
      5 &     & 2.0--2.5 & 0 &  247 & 17 & 264\\
      6 &     & 2.5--3.0 & 0 &  137 & 29 & 166\\
      7 &     & 3.0--3.5 & 0 &   31 & 34 &  65\\
     
     \hline
     8 & 400 & 0.0--0.5 & 45 & 910 & 342&   1297\\
     9 &     & 0.5--1.0 & 31 & 583 & 166&   780\\
     10 &     & 1.0--1.5 & 17 & 340 & 140&  497\\
     11 &     & 1.5--2.0 & 10&  281 & 134&  425\\
     12 &     & 2.0--2.5 & 13 & 162 &  68&  243\\
     13 &     & 2.5--3.0 & 5 &   96 &  71&  172\\
     14 &     & 3.0--3.5 & 0 &   38 &  48&   86\\
    
     \hline
     15 & 500 & 0.0--0.5 & 210 & 643 & 425& 1278\\
     16 &     & 0.5--1.0 & 61 &  434 & 319& 814\\
     17 &     & 1.0--1.5 & 42 &  293 & 219& 554\\
     18 &     & 1.5--2.0 & 22 &  186& 172& 380\\
     19 &     & 2.0--2.5 & 12 &   99& 136&  247\\
     20 &     & 2.5--3.0 & 7  &   53& 95&  155\\
     21 &     & 3.0--3.5 & 18  &  20& 34&  72\\
     
 \hline
     22 & 600 & 0.0--0.5 & 176 & 434 & 712 & 1322\\
     23 &     & 0.5--1.0 & 101 & 283 & 399 & 783\\
     24 &     & 1.0--1.5 & 71 & 150 & 327& 548\\
     25 &     & 1.5--2.0 & 23 & 88 & 257 & 368\\
     26 &     & 2.0--2.5 & 15 & 61 & 172 & 248\\
     27 &     & 2.5--3.0 & 7 &  39 & 125 & 171\\
     28 &     & 3.0--3.5 & 3  & 10  & 47 &  60\\
    
     \hline
    
     \label{tab:tabs2}
    \end{tabular}
  \end{center}
\end{table}

\begin{table}[h!]
  \begin{center}
\caption{List of trajectories from MS-ARMD}
    \begin{tabular}{l|m|j|r|n|q|x}
      \textbf{} & \textbf{$T$[K]} & $b$ & \textbf{Reactive} & \textbf{Complex} & \textbf{Nonreactive}& \textbf{Total}\\
       \hline
      1 & 300 & 0.0--0.5 & 70 & 15& 688& 773\\
      2 &     & 0.5--1.0 & 42 & 6 & 448& 496\\
      3 &     & 1.0--1.5 & 73 & 9 & 729& 811\\
      4 &     & 1.5--2.0 & 95 & 14& 903& 1012\\
      5 &     & 2.0--2.5 & 22 & 5 & 242& 269\\
      6 &     & 2.5--3.0 & 7 &  1 &  98& 106\\
      7 &     & 3.0--3.5 & 6 &  1 &  26&  33\\
      
     \hline
     8 & 400 & 0.0--0.5 & 74 & 14 & 799 & 897\\
     9 &     & 0.5--1.0 & 83 & 24 & 1063& 1170\\
     10&     & 1.0--1.5 & 50 & 9  & 1010& 1069\\
     11&     & 1.5--2.0 & 9 &  1  &  110& 120\\
     12 &     & 2.0--2.5 & 12 & 1  &  105& 118\\
     13 &     & 2.5--3.0 & 7 &  4  &   97&  108\\
     14 &     & 3.0--3.5 & 0 &  1  &    17&  18\\
    
     \hline
     15 & 500 & 0.0--0.5 & 117& 40& 964& 1121\\
     16 &     & 0.5--1.0 & 93 & 24& 844& 961\\
     17 &     & 1.0--1.5 & 43 & 14& 533& 590\\
     18 &     & 1.5--2.0 & 29 & 11& 409& 449\\
     19 &     & 2.0--2.5 & 14 &  5& 208& 227\\
     20 &     & 2.5--3.0 & 2  &  2& 103& 107\\
     21 &     & 3.0--3.5 & 2  &  2&  41& 45\\
     
     \hline
     22 & 600 & 0.0--0.5 & 154 & 30 & 1230& 1414\\
     23 &     & 0.5--1.0 & 114 & 23 &  552&  689\\
     24 &     & 1.0--1.5 & 52  & 8  &  485&  545\\
     25 &     & 1.5--2.0 & 38  & 10 &  468&  516\\
     26 &     & 2.0--2.5 & 23  &  7 &  184&  214\\
     27 &     & 2.5--3.0 & 2   &  2 &   86&   90\\
     28 &     & 3.0--3.5 & 1   &  0 &   31&   32\\

     \hline
     \label{tab:tabs3}
    \end{tabular}
  \end{center}
\end{table}

\end{scriptsize}

%\bibliography{mgoch4}